\documentclass[12pt,a4paper]{article}

\usepackage[left=2.5cm,top=2.5cm,bottom=2.5cm,right=2.5cm]{geometry}
\usepackage{graphicx}
\usepackage{amssymb}
\usepackage{amsthm}
\usepackage{amsmath}
\usepackage{amsfonts}
\usepackage{algorithm, algorithmic}
\usepackage{color}

\usepackage{cite}

\usepackage{enumitem}

\newtheorem{thm}{Theorem}[section]

\newtheorem{lem}[thm]{Lemma}

\theoremstyle{definition}

\theoremstyle{remark}
\newtheorem{rem}{Remark}[section]

\numberwithin{equation}{section}

\begin{document}

\title{Improved in-place associative integer sorting}


\author{A. Emre CETIN \\
email: aemre.cetin@gmail.com}

\maketitle

\begin{abstract}

A novel integer sorting technique was proposed replacing bucket sort, distribution counting sort and address calculation sort family of algorithms which requires only constant amount of additional memory. The technique was inspired from one of the ordinal theories of ``serial order in behavior" and explained by the analogy with the three main stages in the formation and retrieval of memory in cognitive neuroscience namely (i) {\em practicing}, (ii) {\em storing} and (iii) {\em retrieval}. 

In this study, the technique is improved both theoretically and practically and an algorithm is obtained which is faster than the former making it more competitive. With the improved version, $n$ integers $S[0 \ldots n-1]$ each in the range $[0, n-1]$ are sorted exactly in $\mathcal{O}(n)$ time while the complexity of the former technique was the recursion $T(n) = T(\frac{n}{2}) + \mathcal{O}(n)$ yielding $T(n) = \mathcal{O}(n)$.

\end{abstract}


\section{Introduction}\label{sec:intro}


Nervous system is considered to be closely related and described with the ``serial order in behavior" in cognitive neuroscience~\cite{Lashley,Lashley_1} with three basic theories which cover almost all {\em abstract data types} used in computer science. These are~\cite{Henson} chaining theory, positional theory and ordinal theory.

Chaining theory is the extension of stimulus-response (reflex chain) theory, where each response can become the stimulus for the next. From an information processing perspective, comparison based sorting algorithms that sort the lists by making a series of decisions relying on comparing keys can be classified under chaining theory. Each comparison becomes the stimulus for the next. Hence, keys themselves are associated with each other. Some important examples are quick sort~\cite{Hoare}, shell sort~\cite{Shell}, merge sort~\cite{Burnetas} and heap sort~\cite{Williams}.

Positional theory assumes order is stored by associating each element with its position in the sequence. The order is retrieved by using each position to cue its associated element. This is the method by which conventional (Von  Neumann) computers store and retrieve order, through routines accessing separate addresses in memory. Content-based sorting algorithms where decisions rely on the contents of the keys can be classified under this theory. Each key is associated with a position depending on its content. Some important examples are distribution counting sort~\cite{Seward,Feurzig}, address calculation sort~\cite{Isaac,Tarter,Flores,Jones,Gupta,Suraweera}, bucket sort\cite{mahmoud:2000, Cormen} and radix sort~\cite{knuth:vol3,mahmoud:2000,sedgewick:algorithms_in_C, Cormen}.

Ordinal theory assumes order is stored along a single dimension, where that order is defined by relative rather than absolute values on that dimension. Order can be retrieved by moving along the dimension in one or the other direction. This theory need not assume either the item-item nor position-item associations of the previous theories.

One of the ordinal theories of serial order in behavior is that of Shiffrin and Cook\cite{Shiffrin} which suggests a model for short-term forgetting of item and order information of the brain. It assumes associations between elements and a ``node'', but only the nodes are associated with one another. By moving inwards from nodes representing the start and end of the sequence, the associations between nodes allow the order of items to be reconstructed~\cite{Henson}.


The main difficulties of all distributive sorting algorithms is that, when the keys are distributed using a hash function according to their content, several of them may be clustered around a loci, and several may be mapped to the same location. These problems are solved by inherent three basic steps of associative sort~\cite{ecetin} (i) practicing, (ii) storing and (iii) retrieval, which are the three main stages in the formation and retrieval of memory in cognitive neuroscience.

\subsection{Original Technique}\label{subsec:intro1}

As in the ordinal model of Shiffrin and Cook, it is assumed that associations are between the integers in the list space and the nodes in an imaginary linear subspace that spans a predefined range of integers. The imaginary subspace can be defined anywhere on the list space $S[0\ldots n-1]$ provided that its boundaries do not cross over that of the list. The range of the integers spanned by the imaginary subspace is upper bounded by the number of integers $n$ but may be smaller and can be located anywhere making the technique in-place, i.e., beside the input list, only a constant amount of memory locations are used for storing counters and indices. Furthermore, this definition reveals the asymptotic power of the technique with increasing $n$ with respect to the range of integers, as well.

An association between an integer and the imaginary subspace is created by a node using a monotone bijective hash function that maps the integers in the predefined interval to the imaginary subspace. The process of creating a node by mapping a distinct integer to the imaginary subspace is ``practicing a distinct integer of an interval''. Since imaginary subspace is defined on the list space, this is just swapping. Once a node is created, the redundancy due to the association between the integer and the position of the node releases the word allocated to the integer in the physical memory except for one bit which tags the word as a node for interrogation purposes. All the bits of the node except the tag bit can be cleared and used to encode any information. Hence, they are the ``record'' of the node and the information encoded into a record is the ``cue'' by which cognitive neuro-scientists describe the way that the brain recalls the successive items in an order during retrieval. For instance, it will be foreknown from the tag bit that a node has already been created while another occurrence of that particular integer is being practiced providing the opportunity to count other occurrences. The process of counting other occurrences of a particular integer is ``practicing all the integers of an interval'', i.e., rehearsing used by cognitive neuro-scientists to describe the way that the brain manipulates the sequence before storing in a short (or long) term memory. On the other hand, the tag bit discriminates the word as node and the position of the node lets the integer be retrieved back from the imaginary subspace using the inverse hash function.  

Practicing does not need to alter the value of other occurrences, i.e., only the first occurrence is altered while being practiced from where a node is created. All other occurrences of that particular integer remain in the list space but become meaningless. Hence they are ``idle integers''. On the other hand, practicing does not need to alter the position of idle integers as well, unless another distinct integer creates a node exactly at the position of an idle integer while being practiced. In such a case, the idle integer is moved to the former position of the integer that creates the new node. This makes associative sort unstable, i.e., equal integers may not retain their original relative order. However, an imaginary subspace can create other subspaces and associations using the idle integers that were already practiced by manipulating either their position or value or both. Hence, a part of linear algebra and related fields of mathematics can be applied on subspaces to solve such problems. 

Once all the integers in the predefined interval are practiced, the nodes that are dispersed in the imaginary subspace are clustered in a systematic way, i.e., the distance between the nodes are closed to a direction retaining their relative order. This is the {\em storing} phase of associative sort where the received, processed and combined information to construct the sorted permutation of the practiced interval is stored in the short-term memory (for instance, beginning of the list). When the nodes are moved towards a direction, it is not possible to retain the association between the imaginary subspace and list space. However, the record of a node can be further used to encode the absolute position of that node as well, or maybe the relative position or how much that node is moved relative to its absolute or relative position during storing. Unfortunately, this requires that a record is enough to store both the position of the node and the number of idle integers practiced by that node. However, as explained earlier, further associations can be created using the idle integers that were already practiced by manipulating either their position or value or both. Hence, if the record is enough, it can store both the positional information and the number of idle integers. If not, an idle integer can be associated accompanying the node to supply additional space for it for the positional information.
 
Finally, the sorted permutation of the practiced interval is constructed in the list space, using the stored information in the short-term memory. This is the {\em retrieval} phase of associative sort that depends on the information encoded into the record of a node. If the record is enough, it stores both the position of the node and the number of idle integers. If not, an associated idle integer accompanying the node stores the position of the node while the record holds the number of idle integers. The positional information cues the recall of the integer using the inverse hash function. This is ``integer retrieval'' from imaginary subpace. Hence, the retrieved integer can be copied on the list space as much as it occurrs.

Hence, moving through nodes that represent the start and end of practiced integers as well as retaining their relative associations with each other even when their positions are altered by cuing allow the order of integers to be constructed in linear time in-place.

The adjective ``associative'' derived from two facts where the first one mentioned above describes the technique. The second one is that, although associative sort replaces all derivatives of the content based sorting algorithms such as distribution counting sort~\cite{Seward,Feurzig}, address calculation sort~\cite{Isaac,Tarter,Flores,Jones,Gupta,Suraweera} and bucket sort~\cite{mahmoud:2000,Cormen} on a RAM, it seems to be more efficient on a``content addressable memory'' (CAM) known as ``associative memory'' which in one word time find a matching segment in tag portion of the word and reaches the remainder of the word~\cite{Hanlon}. For associative sort developed on a RAM, the nodes of the imaginary subspace (tagged words) and the integers of the list space (untagged words) are processed sequentially which will be a matter of one word time for a CAM to retrieve previous or next tagged or untagged word.

The technique seems to be efficient and applicable for other problems, as well, such as hashing, searching, element distinction, succinct data structures, gaining space, etc. For instance, there are several space gaining techniques available and widely used in the literature for in-place and minimum space algorithms~\cite{Andersson_1,Franceschini_1,Katajainen,Katajainen_1}. However, as known to the author, all these in-place and minimum space algorithms have a dedicated explicit technique that is used only for space gaining purpose. On the contrary, gaining space is an inherent step of associative sort which improves its performance and can be used explicitly.

From complexity point of view, associative sort shows similar characteristics with bucket sort and distribution counting sort. Hence, it can be thought of as {\em in-place associative bucket sort} or {\em in-place associative distribution counting sort}. Distribution counting sort is seldom discussed in the literature although it has been around more than 50 years since proposed by Seward~\cite{Seward} in 1954 and by Feurzig \cite{Feurzig} in 1960, independently, and known to be the method that makes radix sort possible on digital computer. It is known to be very powerful when the integers have small range. Given $n$ integers $S[0\ldots n-1]$ each in the range $[0,m-1]$, its time-complexity is $\mathcal{O}(n+m)$ and requires $n+m$ additional space for a stable and $m$ for an unstable sort. Hence, distribution counting sort becomes efficient and practical when $m=\mathcal{O}(n)$ defining its time-space trade-offs. On the other hand, bucket sort is a generalization of distribution counting sort. In fact, if each bucket has size $1$, then bucket sort degenerates to distribution counting sort. However, the variable bucket size allows it to use $\mathcal{O}(n)$ memory instead of $\mathcal{O}(m+n)$ memory. Its average case time complexity is $\mathcal{O}(n+m)$ and if $m=\mathcal{O}(n)$, then it becomes $\mathcal{O}(n)$. Its worst case time complexity is $\mathcal{O}(n^2)$.

\subsection{Improved Technique}\label{subsec:intro2}

In this study, with a simple revision, the associative sorting technique is improved both theoretically and practically  and a faster technique is achieved. During storing where the nodes are clustered at the beginning of the list retaining their relative order, the positional information ($\log n$ bits) of a node is encoded into either its record or an idle-integer accompanying the node. However, the tag bit discriminates the word as a node in the list space and if ignored during storing it will continue to discriminate the word as a node. This means that, if only the records ($w-1$ bits) of the nodes are clustered at the beginning of the list (short-term memory) retaining their relative order, there will be $n_d$ nodes dispersed in the list space, and $n_d$ records in the short-term memory ($S[0, \ldots n_d-1]$) after storing. Hence, a one-to-one correspondence is obtained with the clustered records and the nodes (tagged words) of the list. Therefore, retrieval phase can search the list from right to left for the first tagged word, retrieve the integer from the imaginary subspace through that node, read its number of occurrence from its record $S[n_d-1]$ in the short-term memory and expand it over the list starting at $S[n_d+n_c-1]$ where $n_c$ is the number of idle integers. Afterwards, the processed tag bit can be cleared and a new search to the left can be carried for the next tagged word which will correspond to the next record $S[n_d-2]$ of the short-term memory. This can continue until all the integers are retrieved from short-term memory resulting in the sorted permutation of the practiced integers.

While the former technique was capable of sorting integers that satisfy $S[i] - \delta + \epsilon < n$ where $\delta=\min(S)$ and $\epsilon \in [0,\frac{n}{2}]$ is defined by Eqn.4.2~\cite{ecetin}, the improved version sorts the integers that satisfy $S[i] - \delta < n$. Hence, $n$ integers $S[0, \ldots n-1]$ each in the range $[0, n-1]$ will be sorted exactly in $\mathcal{O}(n)$ time while the complexity of the former technique was the recursion $T(n) = T(\frac{n}{2}) + \mathcal{O}(n)$ yielding $T(n) = \mathcal{O}(n)$.

With this introductory information, the contribution of this study is,
\begin{description}[leftmargin=0pt]

\item[{\bf A practical sorting algorithm}] that sorts $n$ integers $S[0\ldots n-1]$ each in the range $[0,m-1]$ using $\mathcal{O}(1)$ extra space in $\mathcal{O}(n+m)$ time for the worst, $\mathcal{O}(m)$ time for the average (uniformly distributed integers) and $\mathcal{O}(n)$ time for the best case. The ratio $\frac{m}{n}$ defines the efficiency (time-space trade-offs) of the algorithm letting very larges lists to be sorted in-place. The algorithm is simple and practical replacing bucket sort, distribution counting sort and address calculation sort family of algorithms improving the space requirement to only $\mathcal{O}(1)$ extra words.

Practical comparisons for $1$ million 32 bit integers with quick sort showed that associative sort is roughly $3$ times faster for uniformly distributed integers when $m=n$. When $\frac{m}{n} = 10$ performances are same. When $\frac{m}{n} = \frac{1}{10}$ associative sort becomes roughly $4$ times faster than quick sort. If the distribution is exponential, associative sort shows better performance up to $\frac{m}{n} \approx 25$ when compared with quick sort. 

Practical comparisons for $1$ million 32 bit integers showed that radix sort is $2$ times faster for uniformly distributed integers when $m=n$. However, associative sort is $2$ times faster than radix sort when $\frac{m}{n} = \frac{1}{10}$. Further decreasing the ratio to $\frac{m}{n} =\frac{1}{100}$, associative sort becomes more than $3$ times faster than radix sort. 

Practical comparisons for $1$ million 32 bit integers showed that bucket sort with each bucket of size one (hence distribution counting sort) is $2$ times faster than associative sort for $\frac{m}{n} = 1$. Bucket sort is still slightly better but the performances get closer when $\frac{m}{n} < \frac{1}{10}$ and $\frac{m}{n} > 10$.  

Even omitting its space efficiency for a moment, associative sort asymptotically outperforms all content based sorting algorithms when $n$ is large relative to $m$. 

\end{description}






\section{Definitions}\label{sec:pre}
Given a {\em list} $S$ of $n$ {\em integers}, $S[0], S[1],\ldots , S[n-1]$, the problem is to sort the integers in ascending or descending order. The notations used throughout the study are: 
\begin{enumerate} [label=({\roman{*}}), nosep]
\item Universe of integers is assumed $\mathbb{U} = [ 0 \ldots 2^{w}-1]$ where $w$ is the fixed word length.

\item Maximum and minimum integers of a list are, $\max (S) = \max(a \vert a \in S)$ and $\min (S) = \min(a \vert a \in S)$, respectively. Hence, range of the integers is, $m = \max (S) - \min (S) + 1$.

\item The notation $B \subset A$ is used to indicated that $B$ is a proper subset of $A$.

\item For two lists $S_{1}$ and $S_{2}$, $\max (S_{1}) < \min (S_{2})$ implies $S_{1} < S_{2}$.

\end{enumerate}

\begin{description}[leftmargin = 0pt]

\item[{\bf Universe of integers.}] When an integer is first practiced, a node is created releasing $w$ bits of the integer free. One bit is used to tag the word as a node. Hence, it is reasonable to doubt that the tag bit limits the universe of integers because all the integers should be untagged and in the range $[0,2^{w-1}-1]$ before being practiced. But, we can,
\begin{enumerate}[label=(\roman{*}), itemindent = * , nosep]
\item partition $S$ into $2$ disjoint sublists $S_1 < 2^{w-1} \le S_2$ in $\mathcal{O}(n)$ time with well known in-place partitioning algorithms as well as stably with~\cite{Katajainen},
\item shift all the integers of $S_2$ by $-2^{w-1}$, sort $S_1$ and $S_2$ associatively and shift $S_2$ by $2^{w-1}$.
\end{enumerate}
There are other methods to overcome this problem. For instance, 
\begin{enumerate}[label=(\roman{*}), itemindent = * , nosep]
\item sort the sublist $S[0\ldots (n/ \log n)-1]$ using the optimal in-place merge sort~\cite{Salowe},
\item compress $S[0\ldots (n/ \log n)-1]$ by Lemma~1 of~\cite{Franceschini_1} generating $\Omega(n)$ free bits,
\item sort $S[(n/ \log n)\ldots n-1]$ associatively using $\Omega(n)$ free bits as tag bits,
\item uncompress $S[0\ldots (n/ \log n)-1]$ and merge the two sorted sublists in-place in linear time by~\cite{Salowe}.
\end{enumerate}

\item [{\bf Number of integers.}] If practicing a distinct integer lets us to use $w-1$ bits to practice other occurrences of that integer, we have $w-1$ free bits by which we can count up to $2^{w-1}$ occurrences including the first integer that created the node. Hence, it is reasonable to doubt again that there is another restriction on the size of the lists, i.e., $n \le 2^{w-1}$. But a list can be divided into two parts in $\mathcal{O}(1)$ time and those parts can be merged in-place in linear time by~\cite{Salowe} after sorted associatively.

\end{description}

Hence, for the sake of simplicity, it will be assumed that $n \le 2^{w-1}$ and all the integers are in the range $[0,2^{w-1}-1]$ throughout the study.



\section{Sorting $n$ Integers}\label{subsec:es_multiple}

In this section, the improved associative sorting technique based on the three basic steps namely (i) practicing, (ii) storing and (iii) retrieval will be introduced.

\begin{lem}\label{lem:sorting_seq_of_distinct}
Given $n<=2^{w-1}$ integers $S[0...n-1]$ each in the range $[0,2^{w-1}-1]$, all the integers in the range $[\delta,\delta+n-1]$ where $\delta=\min(S)$ can be sorted associatively at the beginning of the list in $\mathcal{O}(n)$ time using $\mathcal{O}(1)$ constant space.
\end{lem}

\begin{proof}
Given $n<=2^{w-1}$  distinct integers $S[0...n-1]$ each in the range $[0,2^{w-1}-1]$, it is not possible to construct a monotone bijective hash function that maps all the integers of the list into $j \in [0,n-1]$ without additional storage space. At least $\Theta (n + \log w)$ bits are required for a minimal perfect hash function~\cite{Belazzougui} which is not bijective and monotone. However, a monotone bijective hash function can be constructed as a partial function~\cite{rosen:discrete_math_handbook} that assigns each integer of $S_1 \subset S$ in the range $[\delta,\delta+n-1]$ with $\delta=\min(S)$ to exactly one element in $j \in [0,n-1]$. The partial monotone bijective hash function of this form is,
\begin{equation}\label{eqn:hash_func_algo}
\begin{split}
j=S[i]-\delta \quad \text{if} \quad S[i] - \delta < n
\end{split}
\end{equation}

With this definition, the proof has three basic steps of associative sort: 
\begin{enumerate}[label=(\roman{*})]
\item Practice all the integers of the interval $[\delta, \delta+n-1]$ into $Im[0 \ldots n-1]$ over $S[0 \ldots n-1]$.

\item Store only the records ($w-1$ bits) of the nodes at the beginning of the list (short-term memory) retaining their relative order. Hence, a one-to-one correspondence is obtained with the stored records and the nodes (tagged words) of the list. 

\item Retrieve the sorted permutation of the practiced interval by searching the tagged words of the list backwards to retrieve the integers from the imaginary subspace. When an integer is retrieved from the imaginary subspace, read the number of occurrence of that integer from the corresponding record of the short-term memory and expand over the list backwards as many as it occurs.

\end{enumerate}
\end{proof}

\subsection{Practicing Phase}\label{sec:counting_mul}

Details of practicing phase can be found in~\cite{ecetin}.

\begin{enumerate}[label=\bf{Algorithm \Alph{*}.}, ref=Algorithm \Alph{*}, leftmargin=0pt, itemindent=*, start=1] 
\item \label{algorithm:es_fgl_mul} Practice all the integers of the interval $[\delta,\delta+n-1]$ into the imaginary subspace $Im[0\ldots n-1]$ over $S[0 \ldots n-1]$ using Eqn.~\ref{eqn:hash_func_algo}. It is assumed that the minimum of the list $\delta=\min(S)$ is known.
\end{enumerate}

\begin{enumerate}[label=\bf{A\arabic{*}.}, ref=A\arabic{*}, itemindent=* , nosep]
\item initialize $i = 0$;\label{algo1:item0}
\item if MSB of $S[i]$ is $1$, then $S[i]$ is a node. Hence, increase $i$ and repeat this step;\label{algo1:item2}
\item if $S[i] - \delta \ge n$, then $S[i]$ is an integer of $S_2$ that is out of the practiced interval. Hence, increase $n_d'$ that counts the number of integers of $S_2$, update $\delta'=min(\delta', S[i])$, increase $i$ and goto to step \ref{algo1:item2};\label{algo1:item3}
\item otherwise, $S[i]$ is an integer to be practiced. Hence, calculate $j = S[i] - \delta$ (Eqn.~\ref{eqn:hash_func_algo});\label{algo1:item4}
\item if MSB of $S[j]$ is $0$, then $S[i]$ is the first integer that will create the node at $j$. Move $S[j]$ to $S[i]$, clear $S[j]$ and set its MSB to $1$ making it a node. If $j \le i$ increase $i$. Increase $n_d$ that counts the number of distinct integers (nodes), and goto step \ref{algo1:item2}; \label{algo1:item5}
\item otherwise, $S[j]$ is a node that has already been created. Hence, clear MSB of $S[j]$, increase $S[j]$ (number of idle integers) and set its MSB back to $1$. Increase $i$ and $n_c$ that counts the number of total idle integers over all distinct integers and goto step \ref{algo1:item2};
\end{enumerate}

\subsection{Storing Phase}\label{sec:partitioning_mul}

Practicing creates $n_d$ nodes and $n_c$ idle integers. This means $n_d$ integers of $S_1$ are mapped into the imaginary subspace creating nodes that are dispersed with relative order in $Im[0 \ldots n-1]$ over $S[0 \ldots n-1]$ depending on the statistical distribution of the integers. On the other hand, $n_c$ idle integers of $S_1$ are distributed disorderly together with $n_d'$ integers of $S_2$ in the list space.

In storing phase, the records are clustered in a systematic way, i.e., the distance between the records of the imaginary subspace are closed to a direction (beginning of the list) without altering their relative order with respect to each other. As long as the position of the tag bits are not altered, the association between the imaginary subspace and the list space is retained. 

\begin{enumerate}[label=\bf{Algorithm \Alph{*}.}, ref=Algorithm \Alph{*}, leftmargin=0pt, itemindent=*, start=2] 
\item \label{algorithm:es_fgp_mul} Store the records of the practiced interval in the short term memory.  
\end{enumerate}
\begin{enumerate}[label=\bf{B\arabic{*}.}, ref=B\arabic{*}, itemindent=* , nosep]
\item initialize $i = 0$, $j = 0$, $k = n_d$;
\item if MSB of $S[i]$ is $0$, then $S[i]$ is either an idle integer or an integer of $S_2$ that is out of the practiced interval. Hence, increase $i$ and repeat this step; \label{algo2:item1}
\item otherwise, $S[i]$ is a node. Hence, swap least significant $w-1$ bits of $S[i]$ with least significant $w-1$ bits of $S[j]$. Increase $i$ and $j$ and decrease $k$. If $k = 0$ exit, otherwise goto step \ref{algo2:item1}; \label{algo2:item2}
\end{enumerate}


\subsection{Retrieval Phase}\label{sec:decoding_mul}

Storing clusters $n_d$ records of the nodes at $S[0 \ldots n_d-1]$. Hence, $S[0 \ldots n_d-1]$ can be though of as a short-term memory where the encoded information of the $n_d+n_c$ integers of the practiced interval is stored. The stored information is the number of occurrences of the distinct integers. However, the tagged words of the list have a one-to-one correspondence with these $n_d$ records from left to right or vice versa. 

In retrieval phase, the stored information is retrieved from the short term memory $S[0 \ldots n_d-1]$ to construct the sorted permutation of the practiced interval. The short term memory encodes $n_d+n_c$ integers of $S_1$ with $n_d$ permanent records. Each record stores the number of occurrence of a particular integer. On the other hand, the particular integer can be retrieved back to list space through its corresponding node. It is important to note that, if the number of occurrences of a particular integer is $n_i$, then there are $n_i-1$ idle integers in the list. But the record itself represents the integer that is mapped into the imaginary subspace. Hence, it is immediate from this definition that the list can be searched from right to left backwards for the first tagged word to retrieve the integer back to list space (using the inverse hash function) and the integer can be copied as many as it occurs which can be read from the record $S[n_d-1]$ of the node in the short term memory $S[0 \ldots n_d-1]$. Afterwards, the processed tag bit can be cleared and a new search to the right can be carried for the next tagged word which will correspond to the next record $S[n_d-2]$ of the short-term memory. This can continue until all the integers are retrieved resulting in the sorted permutation of the practiced integers. 

It should be noted that, $n_c$ idle integers of $S_1$ and $n_d'$ integers of $S_2$ are distributed disorderly together at $S[n_d \ldots n-1]$. Hence, before proceeding, $n_c$ idle integers should be clustered at the beginning of $S[n_d \ldots n-1]$. Afterwards, the retrieval phase can begin. This is a simple partitioning problem where the tag bits should be taken under care.

\begin{enumerate}[label=\bf{Algorithm \Alph{*}.}, ref=Algorithm \Alph{*}, leftmargin=0pt, itemindent=*, start=3] 
\item \label{algorithm:es_fgpp_mul} Partition $S[n_d \ldots n-1]$ to cluster $n_c$ idle integers to the beginning. 
\end{enumerate}
\begin{enumerate}[label=\bf{C\arabic{*}.}, ref=C\arabic{*}, itemindent=* , nosep]
\item initialize $i = n_d$, $j = n_d$, $k = n_c$;
\item Read $w-1$ bits of $S[i]$ into $s$. If $s - \delta >= n$, then $S[i]$ is an integer of $S_2$ that is out of the practiced interval. Hence, increase $i$ and repeat this step; \label{algo3:item1}
\item otherwise, $S[i]$ is an idle integer. Hence, swap least significant $w-1$ bits of $S[i]$ with least significant $w-1$ bits of $S[j]$. Increase $i$ and $j$ and decrease $k$. If $k = 0$ exit, otherwise goto step \ref{algo3:item1}; \label{algo3:item2}
\end{enumerate}

\begin{enumerate}[label=\bf{Algorithm \Alph{*}.}, ref=Algorithm \Alph{*}, leftmargin=0pt, itemindent=*, start=4] 
\item \label{algorithm:es_fgd_mul} Process the list from right to left to find a node, retrieve the integer from the imaginary subspace through that node, read the number of occurrence of that integer from the corresponding record $S[n_d-1]$ and expand the integer over $S[0 \ldots n_d+n_c-1]$ sequentially right to left backwards. Then find the next node which corresponds to the record at $S[n_d-2]$. Continue until all the integers are retrieved and expanded over $S[0 \ldots n_d+n_c-1]$.
\end{enumerate}
\begin{enumerate}[label=\bf{D\arabic{*}.}, ref=D\arabic{*}, itemindent=* , nosep]
\item initialize $i = n-1$, $j = n_d-1$ and $k=n_d+n_c$,
\item if MSB of $S[i]$ is $0$, then it is not a node. Hence, decrease $i$ and repeat this step;\label{algo4:item0}
\item otherwise, $S[i]$ is a node. Hence, the position $i$ of the node cues the recall of the integer using the inverse of Eqn.~\ref{eqn:hash_func_algo}. Get the number of occurrence of the retrieved integer from its record ($w-1$ bits of $S[j]$) and copy the integer as many as it occurs starting with $w-1$ bits of $S[k-1]$ and decreasing $k$ one for each copied integer. At the end, clear MSB of $S[i]$, decrease $i$, $j$ and $k$. If $k=0$, then exit. Otherwise goto step \ref{algo4:item0};
\end{enumerate}

\begin{description}[leftmargin = 0pt]
\item [{\bf Sequential Version}] After retrieval phase, $n_d'$ integers of $S_2$ are clustered disorderly at $S[n_d+n_c \ldots n-1]$. Hence, the same algorithm can be called to sort this new list $S[n_d' \ldots n-1]$ with its minimum $\delta'$ found at step \ref{algo1:item3}. 

\begin{rem}
Improved associative sort technique is on-line in the sense that after each retrieval phase (\ref{algorithm:es_fgd_mul}), $n_d+n_c$ integers are added to the sorted permutation at the beginning of the list and ready to be used.
\end{rem}

\begin{rem}
Recursive version is not possible as long as the tag bits are not clustered during storing phase.
\end{rem}

\end{description}

\begin{description}[leftmargin = 0pt]

\item [{\bf Complexity}] of the algorithm depends on the range and the number of integers. In each iteration the algorithm is capable of sorting the integers that satisfy $S[i] - \delta < n$. Hence, given uniformly distributed $n$ integers $S[0 \ldots n-1]$ each in the range $[0,n-1]$, the complexity is $\mathcal{O}(n)$.

\item[{\bf Best Case Complexity.}] Given $n$ integers $S[0 \ldots n-1]$, if $n-1$ of them satisfy $S[i] - \delta < n$, then these are sorted in $\mathcal{O}(n)$ time. In the next step, there is one integer left which implies sorting is finished. As a result, time complexity of the algorithm is lower bounded by $\Omega(n)$ in the best case.

\item [{\bf Worst Case Complexity.}] Given $n$ integers $S[0 \ldots n-1]$ and $m=\beta n$, if there is only $1$ integer available in practiced interval at each iteration until the last, in any $j$th step, the only integer $s$ that will be sorted satisfies $s < jn-(j-1)$ which implies that the last alone integer satisfies $s < jn-(j-1) \le \beta n$ from where we can calculate $j$ by $j \le \frac{\beta n-1 }{n-1}$.
In this case, the time complexity of the algorithm is,
\begin{equation}
\mathcal{O}(n) + \mathcal{O}(n-1) + \dotsc + \mathcal{O}(n-j) = (j+1) \mathcal{O}(n) -\mathcal{O}(j^2) < (\beta+1) \mathcal{O}(n)
\end{equation}

Therefore, the algorithm is upper bonded by $(\beta+1) \mathcal{O}(n) = \mathcal{O}(m+n)$ in worst case.

\item[{\bf Average Case Complexity.}] Given $n$ integers $S[0 \ldots n-1]$, if $m = \beta n$ and the integers are uniformly distributed, this means that $\frac{n}{\beta}$ integers satisfy $S[i] < n$. Therefore, the algorithm is capable of sorting $ \frac{n}{\beta}$ integers in $\mathcal{O}(n)$ time during first pass. This will continue until all the integers are sorted. The sum of sorted integers in each iteration can be represented with the series,
\begin{equation} \label{eqn:series_sorted_case3}
\frac{n}{\beta}+\frac{n(\beta-1)}{\beta^2}+\dotsc+\frac{n(\beta-1)^{k-1}}{\beta^{k}}+\dotsc
\end{equation}

It is reasonable to think that the sorting ends when one term is left which means the sum of $k$ terms of this series is equal to $n-1$, from where we can calculate the number of iteration or dept of recursion $k$ which is valid when $\beta > 1$ by,
\begin{equation} \label{eqn:series_sorted_case3_4}
\frac{1}{n} = \frac{(\beta-1)^{k-1}}{\beta^{k}}
\end{equation}
It is seen from Eqn.~\ref{eqn:series_sorted_case3_4} that when $m = 2n$, i.e., $\beta=2$, number of iteration or dept of recursion becomes $k=\log{n}$ and the complexity is the recursion $T(n) = T(\frac{n}{2}) + \mathcal{O}(n)$ yielding $T(n) = \mathcal{O}(n)$. It is known that each step takes $\mathcal{O}(n)$ time. Therefore, the time complexity of the algorithm is,
\begin{equation}\label{eqn:series_complexity_case3}
\begin{split}
\mathcal{O}(n)+\mathcal{O}\bigl(\frac{n(\beta-1)}{\beta}\bigr) +\dotsc +\mathcal{O}\bigl( \frac{n(\beta-1)^{k-1}}{\beta^{k-1}} \bigr)
\end{split}
\end{equation}
from where we can obtain by defining $x= \frac{(\beta-1)}{\beta}$,
\begin{equation}\label{eqn:ud_6}
\mathcal{O}(n) \bigl( 1 +  x + x^2 + x^3 + \cdots + x^{k-1} \bigr) = \mathcal{O}(n) (\frac{1}{1-x} - \frac{x^{k-1}}{1-x}) < \beta \mathcal{O}(n)
\end{equation}
which means that the algorithm is upper bounded by $ \beta \mathcal{O}(n)$ or $\mathcal{O}(m)$ in the average case.

\end{description}

\section{Conclusions}
\label{chap:summaryandconclusion}

In this study, in-place associative integer sorting technique is improved. The improved version can only run sequentially. Recursive version is not possible. While the original technique was capable of sorting integers that satisfy $S[i] - \delta + \epsilon < n$ where $\delta=\min(S)$ and $\epsilon \in [0,\frac{n}{2}]$, the improved version sorts the integers that satisfy $S[i] - \delta < n$. Hence, given $n$ integers $S[0, \ldots n-1]$ each in the range $[0, n-1]$ will be sorted exactly in $\mathcal{O}(n)$ time while the complexity of the former technique was the recursion $T(n) = T(\frac{n}{2}) + \mathcal{O}(n)$ yielding $T(n) = \mathcal{O}(n)$.

Using the technique, the main difficulties of distributive sorting algorithms are solved by its inherent three basic steps: (i) practicing, (ii) storing and (iii) retrieval which are three main stages in the formation and retrieval of memory in cognitive neuroscience. The technique is very simple and straightforward and around 30 lines of C code is enough. 

The technique sorts the integers using only $\mathcal{O}(1)$ extra space in $\mathcal{O}(n+m)$ time for the worst, $\mathcal{O}(m)$ time for the average (uniformly distributed integers) and $\mathcal{O}(n)$ time for the best case. It shows similar characteristics with bucket sort and distribution counting sort. However, it is time-space efficient than both. The ratio $\frac{m}{n}$ defines the efficiency (time-space trade-offs) letting very large lists to be sorted in-place. Furthermore, the dependency of the efficiency on the distribution of the integers is $\mathcal{O}(n)$ which means it replaces all the methods based on address calculation, that are known to be very efficient when the integers have known (usually uniform) distribution and require additional space more or less proportional to $n$. Hence, associative sort asymptotically outperforms all content based sorting algorithms. 

The only drawback of the algorithm is that it is unstable. But, an imaginary subspace can create other subspaces and associations using the idle integers that were already practiced by manipulating either their position or value or both. Hence, different techniques can be developed on subspaces to solve other problems such as stability.

\end{document}